\documentstyle[aps, preprint]{revtex}
\def\beq{\begin{eqnarray}}
\def\eed{\end{eqnarray}}
\begin{document}
\draft

\title{Pseudogap effects on the c-axis charge dynamics in copper oxide
materials}
\author{Shiping Feng$^{1,2,3}$, Feng Yuan$^{2}$, and Weiqiang Yu$^{2}$}
\address{
$^{1}$CCAST (World Laboratory) P. O. Box 8730, Beijing 100080,
China \\
$^{*2}$Department of Physics, Beijing Normal University, Beijing
100875, China \\
$^{3}$National Laboratory of Superconductivity, Academia Sinica,
Beijing 100080, China \\
}
\maketitle
\begin{abstract}
The c-axis charge dynamics of copper oxide materials in the
underdoped and optimally doped regimes has been studied by
considering the incoherent interlayer hopping. It is shown that
the c-axis charge dynamics for the chain copper oxide materials is
mainly governed by the scattering from the in-plane fluctuation,
and the c-axis charge dynamics for the no-chain copper oxide
materials is dominated by the scattering from the in-plane
fluctuation incorporating with the interlayer disorder, which
would be suppressed when the holon pseudogap opens at low
temperatures and lower doping levels, leading to the crossovers
to the semiconducting-like range in the c-axis resistivity and
the temperature linear to the nonlinear range in the in-plane
resistivity.
\end{abstract}

\section{Introduction}

It has become clear in the past several years that copper oxide materials
are among the most complex systems studied in condensed matter physics,
and show many unusual normal-state properties. The complications arise
mainly from (1) strong anisotropy in the properties parallel and
perpendicular to the CuO$_{2}$ planes which are the key structural
element in the whole copper oxide superconducting materials, and (2)
extreme sensitivity of the properties to the compositions (stoichiometry)
which control the carrier density in the CuO$_{2}$ plane \cite{n1}, while
the unusual normal-state feature is then closely related to the fact
that these copper oxide materials are doped Mott insulators, obtained
by chemically adding charge carriers to a strongly correlated
antiferromagnetic (AF) insulating state, therefore the physical
properties of these systems mainly depend on the extent of dopings, and
the regimes have been classified into the underdoped, optimally doped,
and overdoped, respectively \cite{n2}. The normal-state properties of
copper oxide materials in the underdoped and optimally doped regimes
exhibit a number of anomalous properties in the sense that they do not
fit in the conventional Fermi-liquid theory \cite{n2,n3}, and the
mechanism for the superconductivity in copper oxide materials has been
widely recognized to be closely associated with the anisotropic
normal-state properties \cite{n4,n5}. Among the striking features of
the normal-state properties in the underdoped and optimally doped
regimes, the physical quantity which most evidently displays the
anisotropic property in copper oxide materials is the charge
dynamics \cite{n5}, which is manifested by the optical conductivity
and resistivity. It has been show from the experiments that the
in-plane charge dynamics is rather universal within the whole copper
oxide materials \cite{n2,n5}. The in-plane optical conductivity for
the same doping is nearly materials independent both in the magnitude
and energy dependence, and shows the non-Drude behavior at low energies
and anomalous midinfrared band in the charge-transfer gap, while the
in-plane resistivity $\rho_{ab}(T)$ exhibits a linear behavior in the
temperature in the optimally doped regime and a nearly temperature
linear dependence with deviations at low temperatures in the
underdoped regime \cite{n2,n5}. By contrast, the magnitude of the
c-axis charge dynamics in the underdoped and optimally doped regimes
is strongly materials dependent, {\it i.e.}, it is dependent on the
species of the building blocks in between the CuO$_{2}$ planes
\cite{n6,n7,n8,n9,n10}. In the underdoped and optimally doped regimes,
the experimental results \cite{n6,n7,n8,n9,n10} show that the ratio
$R=\rho_{c}(T)/\rho_{ab}(T)$ ranges from $R\sim 100$ to $R >10^{5}$,
this large magnitude of the resistivity anisotropy reflects that the
c-axis mean free path is shorter than the interlayer distance, and
the carriers are tightly confined to the CuO$_{2}$ planes, and also
is the evidence of the incoherent charge dynamics in the c-axis
direction. For the copper oxide materials without the Cu-O chains
in between the CuO$_{2}$ planes \cite{n9}, such as
La$_{2-x}$Sr$_{x}$CuO$_{4}$ systems, the transferred weight in the
c-axis conductivity forms a band peaked at high energy $\omega
\sim 2ev$, and the low-energy spectral weight is quite small and
spread over a wide energy range instead of forming a peak at low
energies, in this case the behavior of the c-axis temperature
dependent resistivity $\rho_{c}(T)$ is characterized by a crossover
from the high temperature metallic-like to the low temperature
semiconducting-like \cite{n9}. However, for these copper oxide
materials with the Cu-O chains in between the CuO$_{2}$ planes
\cite{n10}, such as YBa$_{2}$Cu$_{3}$O$_{7}$ systems, the c-axis
conductivity exhibits the non-Drude behavior at low energies and
weak midinfrared band, moreover, this weak midinfrared band rapidly
decrease with reducing dopings or increasing temperatures, while the
c-axis resistivity $\rho_{c}(T)$ is linear in temperatures in the
optimally doped regime, and shows a crossover from the high
temperature metallic-like behavior to the low temperature
semiconducting-like behavior in the underdoped regime \cite{n10}.
Therefore there are some subtle differences between the chain and
no-chain copper oxide materials.

The c-axis charge dynamics of copper oxide materials has been addressed
from several theoretical viewpoints \cite{n11,n12,n13,n14,n15}. Based
on the concept of dynamical dephasing, Leggett \cite{n11} thus proposed
that the c-axis conduction has to do with scatterings from in-plane
thermal fluctuations, and depends on the ratio of the interlayer
hopping rate of CuO$_{2}$ sheets to the thermal energy. While the
theory of tunneling c-axis conductivity in the incoherent regime has
been given by many researchers \cite{n12}. Based on a highly
anisotropic Fermi-liquid, some effect from the interlayer static
disorder or dynamical one has been discussed \cite{n13}. The similar
incoherent conductivity in the coupled fermion chains has been in more
detail studied by many authors within the framework of the
non-Fermi-liquid theory \cite{n14}. Moreover, the most reliable
result for the c-axis charge dynamics from the model relevant to
copper oxide materials has been obtained by the numerical
simulation \cite{n15}. It has been argued that the in-plane
resistivity deviates from the temperature linear behavior and
temperature coefficient of the c-axis resistivity change sign,
showing semiconducting-like behavior at low temperatures are
associated with the effect of the pseudogap \cite{n9,n10}. To shed
light on this issue, we, in this paper, apply the fermion-spin
approach \cite{n16,n17} to study the c-axis charge dynamics by
considering the interlayer coupling.

The paper is organized as follows. The theoretical framework is
presented in Sec. II. In the case of the incoherent interlayer
hopping, the c-axis current-current correlation function (then the
c-axis optical conductivity) is calculated in terms of the in-plane
single-particle spectral function by using standard formalisms for
the tunneling in metal-insulator-metal junctions \cite{n18,n19}.
Within this theoretical framework, we discuss the c-axis charge
dynamics of the chain copper oxide materials in Sec. III. It is
shown that the c-axis charge dynamics of the chain copper oxide
materials is mainly governed by the scattering from in-plane charged
holons due to spinon fluctuations, and the behavior of the c-axis
resistivity is the metallic-like in the optimally doped regime and
the semiconducting-like in the underdoped regime at low temperatures.
In Sec. IV, the c-axis charge dynamics of the no-chain copper oxide
materials is discussed. Our result shows that the scattering from
the in-plane fluctuation incorporating with the interlayer disorder
dominates the c-axis charge dynamics for the no-chain copper oxide
materials. In this case, the c-axis resistivity exhibits the
semiconducting-like behavior in the underdoped and optimally doped
regimes at low temperatures. Sec. V is devoted to a summary and
discussions. Our results also show that the crossover to the
semiconducting-like range in $\rho_{c}(T)$ is obviously linked with
the crossover from the temperature linear to the nonlinear range in
$\rho_{ab}(T)$, and the common origin for these crossovers is due to
the existence of the holon pseudogap at low temperatures and lower
doping levels.

\section{Theoretical framework}

Among the microscopic models the most simplest for the discussion
of doped Mott insulators is the $t$-$J$ model \cite{n20}, which is
originally introduced as an effective Hamiltonian of the Hubbard
model in the strong coupling regime, where the electron become
strongly correlated to avoid the double occupancy. The interest in
the $t$-$J$ model is stimulated by many researchers' suggestions
that it may contain the essential physics of copper oxide materials
\cite{n20,n21}. On the other hand, there is a lot of evidence from
the experiments and numerical simulations in favour of the $t$-$J$
model as the basic underlying microscopic model \cite{n3,n22}.
Within each CuO$_{2}$ plane, the physics property may be described
by the two-dimensional (2D) $t$-$J$ model,
\begin{eqnarray}
H_{l}=-t\sum_{i\hat{\eta}\sigma}C^{\dagger}_{li\sigma}
C_{li+\hat{\eta}\sigma} + h.c. - \mu \sum_{i\sigma}
C^{\dagger}_{li\sigma}C_{li\sigma} +
J\sum_{i\hat{\eta}}{\bf S}_{li}\cdot {\bf S}_{li+\hat{\eta}} ,
\end{eqnarray}
supplemented by the on-site local constraint $\sum_{\sigma}
C^{\dagger}_{li\sigma}C_{li\sigma}\leq 1$ to avoid the double
occupancy, where $\hat{\eta}=\pm a_{0}\hat{x},\pm a_{0}\hat{y}$,
$a_{0}$ is the lattice constant of the square planar lattice,
which is set as the unit hereafter, $i$ refers to planar sites
within the l-th CuO$_{2}$ plane, $C^{\dagger}_{li\sigma}$
($C_{li\sigma}$) are the electron creation (annihilation) operators,
${\bf S}_{li}=C^{\dagger}_{li}{\bf \sigma}C_{li}/2$ are the spin
operators with ${\bf \sigma}=(\sigma_{x},\sigma_{y},\sigma_{z})$
as the Pauli matrices, and $\mu$ is the chemical potential.
Then the hopping between CuO$_{2}$ planes is considered as
\cite{n15}
\begin{eqnarray}
H=-t_{c}\sum_{l\hat{\eta}_{c}i\sigma}C^{\dagger}_{li\sigma}
C_{l+\hat{\eta}_{c}i\sigma} + h.c. +\sum_{l}H_{l},
\end{eqnarray}
where $\hat{\eta}_{c}=\pm c_{0}\hat{z}$, $c_{0}$ is the interlayer
distance, and has been determined from the experiments \cite{n23}
as $c_{0} > 2a_{0}$. As mentioned above, the experimental results
show that the c-axis charge dynamics in the underdoped and optimally
doped regimes is incoherent, therefore the c-axis momentum can not
be defined \cite{n24}. Moreover, the absence of the coherent c-axis
charge dynamics is a consequence of the weak interlayer hopping
matrix element $t_{c}$, but also of a strong intralayer scattering,
{\it i.e.}, $t_{c}\ll t$, and therefore the common CuO$_{2}$ planes
in copper oxide materials clearly dominate the most normal-state
properties. In this case, the most relevant for the study of the
c-axis charge dynamics is the results on the in-plane conductivity
$\sigma_{ab}(\omega)$ and related single-particle spectral function
$A(k,\omega)$.

Since the strong electron correlation in the $t$-$J$ model manifests
itself by the electron single occupancy on-site local constraint,
then the crucial requirement is to impose this electron on-site local
constraint for a proper understanding of the physics of copper oxide
materials.  To incorporate this local constraint, the fermion-spin
theory based on the charge-spin separation has been proposed
\cite{n16,n17}. According to the fermion-spin theory, the constrained
electron operators in the $t$-$J$ model is decomposed as \cite{n16},
\begin{eqnarray}
C_{li\uparrow}=h^{\dagger}_{li}S^{-}_{li}, ~~~~~~ C_{li\downarrow}
=h^{\dagger}_{li}S^{+}_{li},
\end{eqnarray}
with the spinless fermion operator $h_{i}$ keeps track of the
charge (holon), while the pseudospin operator $S_{i}$ keeps track
of the spin (spinon). The main advantage of this approach is
that the electron on-site local constraint can be treated exactly
in analytical calculations. In this case, the low-energy behavior
of the $t$-$J$ model (2) in the fermion-spin representation can be
written as \cite{n25},
\begin{mathletters}
\begin{eqnarray}
H &=& t_{c}\sum_{l\hat{\eta}_{c}i}h^{\dagger}_{l+\hat{\eta}_{c}i}
h_{li}(S^{+}_{li}S^{-}_{l+\hat{\eta}_{c}i}+S^{-}_{li}
S^{+}_{l+\hat{\eta}_{c}i})+ \sum_{l}H_{l}, ~~~~~~~~~~~~~~~~\\
H_{l} &=& t\sum_{i\hat{\eta}}h^{\dagger}_{li+\hat{\eta}}h_{li}
(S^{+}_{li}S^{-}_{li+\hat{\eta}}+S^{-}_{li}S^{+}_{li+\hat{\eta}})
+ \mu \sum_{i}h^{\dagger}_{li}h_{li} + J_{eff}\sum_{i\hat{\eta}}
({\bf S}_{li}\cdot {\bf S}_{li+\hat{\eta}}),
\end{eqnarray}
\end{mathletters}
where $J_{eff}=J[(1-\delta)^{2}-\phi ^{2}]$, the in-plane holon
particle-hole parameter $\phi=\langle h^{\dagger}_{li}
h_{li+\hat{\eta}}\rangle$, and $S^{+}_{li}$ and $S^{-}_{li}$ are
the pseudospin raising and lowering operators, respectively. As
a consequence, the kinetic part in the $t$-$J$ model has been
expressed as the holon-spinon interaction in the fermion-spin
representation, which dominates the charge and spin dynamics in
copper oxide materials in the underdoped and optimally doped
regimes. The spinon and holon may be separated at the mean-field
level, but they are strongly coupled beyond mean-field
approximation (MFA) due to fluctuations.

The mean-field theory within the fermion-spin formalism in the
underdoped and optimally doped regimes without AF long-range-order
(AFLRO) has been developed \cite{n17}, and the in-plane mean-field
spinon and holon Green's functions $D^{(0)}_{ab}(i-j,\tau-\tau')=-
\langle T_\tau S^{+}_{li}(\tau)S^{-}_{lj}(\tau')\rangle_0$ and
$g^{(0)}_{ab}(i-j,\tau-\tau')=-\langle T_\tau h_{li}(\tau)
h^{\dagger}_{lj}(\tau')\rangle_0$ have been evaluated \cite{n17} as,
\begin{eqnarray}
D^{(0)}_{ab}({\bf k},\omega)={B_{k}\over 2\omega (k)}\left (
{1\over \omega -\omega (k)}-{1\over \omega +\omega (k)}\right ),
\end{eqnarray}
\begin{eqnarray}
g^{(0)}_{ab}({\bf k},\omega)={1\over \omega-\xi_k},
\end{eqnarray}
respectively, where $B_{k}=\lambda [(2\epsilon\chi_{z}+\chi)
\gamma_{k}-(\epsilon\chi+2\chi_{z})]$, $\gamma_{{\bf k}}=(1/Z)
\sum_{\eta}e^{i{\bf k}\cdot\hat{\eta}}$, $\lambda =2ZJ_{eff}$,
$\epsilon=1+2t\phi/J_{eff}$, $Z$ is the number of the nearest
neighbor sites at the plane, while the in-plane mean-field spinon
spectrum
\begin{eqnarray}
\omega^{2}(k)=\lambda^{2}\left (\alpha\epsilon[\chi_{z}\gamma_{k}
+{1\over 2Z}\chi ]-[\alpha C_{z}+{1\over 4Z}(1-\alpha)]\right )
(\epsilon\gamma_{k}-1) \nonumber \\
+\lambda^{2}\left (\alpha\epsilon[{1\over 2}\chi\gamma_{k}+
{1\over Z}\chi_{z}]-{1\over 2}\epsilon[\alpha C+{1\over 2Z}
(1-\alpha)]\right )(\gamma_{k}-\epsilon),
\end{eqnarray}
and the in-plane mean-field holon spectrum
$\xi_k=2Z\chi t\gamma_k+\mu$, with the in-plane spinon correlation
functions $\chi=\langle S_{li}^{+}S_{li+\hat{\eta}}^{-}\rangle$,
$\chi_{z}=\langle S_{li}^{z}S_{li+\hat{\eta}}^{z}\rangle$,
$C=(1/Z^{2})\sum_{\hat{\eta},\hat{\eta'}}\langle
S_{li+\hat{\eta}}^{+}S_{li+\hat{\eta'}}^{-}\rangle$, and
$C_{z}=(1/Z^{2})\sum_{\hat{\eta}, \hat{\eta'}}\langle
S_{li+\hat{\eta}}^{z}S_{li+\hat{\eta'}}^{z}\rangle$. In order not
to violate the sum rule of the correlation function $\langle
S^{+}_{li}S^{-}_{li}\rangle=1/2$ in the case without AFLRO, the
important decoupling parameter $\alpha$ has been introduced in the
mean-field calculation \cite{n17}, which can be regarded as the
vertex correction. The mean-field order parameter $\chi$, $C$,
$\chi_z$, $C_z$, $\phi$ and chemical potential $\mu$ have been
determined \cite{n17} by the self-consistent equations.

Within the 2D $t$-$J$ model, the in-plane charge dynamics in copper
oxide materials has been discussed \cite{n25} by considering
fluctuations around this mean-field solution, and the result
exhibits a behavior similar to that seen in the experiments
\cite{n2,n3} and numerical simulations \cite{n26}. In the framework
of the charge-spin separation, an electron is represented by the
product of a holon and a spinon, then the external field can only
be coupled to one of them. According to the Ioffe-Larkin combination
rule \cite{n27,n28}, the physical c-axis conductivity $\sigma_{c}
(\omega)$ is given by,
\begin{eqnarray}
\sigma^{-1}_{c}(\omega)=\sigma^{(h)-1}_{c}(\omega)
+\sigma^{(s)-1}_{c}(\omega),
\end{eqnarray}
where $\sigma^{(h)}_{c}(\omega)$ and $\sigma^{(s)}_{c}(\omega)$
are the contributions to the c-axis conductivity from holons and
spinons, respectively, and can be expressed \cite{n18} as,
\begin{eqnarray}
\sigma^{(h)}_{c}(\omega)=-{{\rm Im}\Pi^{(h)}_{c}(\omega)
\over \omega},~~~~
\sigma^{(s)}_{c}(\omega)=-{{\rm Im}\Pi^{(s)}_{c}(\omega)
\over \omega},
\end{eqnarray}
with $\Pi^{(h)}_{c}(\omega)$ and $\Pi^{(s)}_{c}(\omega)$ are the
holon and spinon c-axis current-current correlation function,
respectively, which are defined as,
\begin{eqnarray}
\Pi^{(s)}_{c}(\tau-\tau')=-\langle T_\tau j^{(s)}_{c}(\tau)
j^{(s)}_{c}(\tau')\rangle,~~
\Pi^{(h)}_{c}(\tau-\tau')=-\langle T_\tau j^{(h)}_{c}(\tau)
j^{(h)}_{c}(\tau')\rangle.
\end{eqnarray}
Within the Hamiltonian (4), the c-axis current densities of spinons
and holons are obtained by the time derivation of the polarization
operator using Heisenberg's equation of motion as,
$j^{(h)}_{c}=2\tilde{t}_{c}e\chi\sum_{l\hat{\eta}_{c}i}
\hat{\eta}_{c}h_{l+\hat{\eta}_{c}i}^{\dagger} h_{li}$ and
$j^{(s)}_{c}=t_{c}e\phi_{c}\sum_{l\hat{\eta}_{c}i}\hat{\eta}_{c}
(S^{+}_{li}S^{-}_{l+\hat{\eta}_{c}i}+S^{-}_{li}
S^{+}_{l+\hat{\eta}_{c}i})$, respectively, with $\tilde{t}_{c}=
t_{c}\chi_{c}/\chi$ is the effective interlayer holon hopping
matrix element, and the order parameters $\chi_{c}$ and $\phi_{c}$
are defined \cite{n17} as $\chi_{c}=\langle S_{li}^{+}
S_{l+\hat{\eta}_{c}i}^{-}\rangle$, and $\phi_{c}=\langle
h^{\dagger}_{li}h_{l+\hat{\eta}_{c}i}\rangle$, respectively. As in
the previous discussions \cite{n25}, a formal calculation for the
spinon part shows that there is no the direct contribution to the
current-current correlation from spinons, but the strongly
correlation between holons and spinons is considered through the
spinon's order parameters entering in the holon part of the
contribution to the current-current correlation, therefore the
charge dynamics in copper oxide materials is mainly caused by
charged holons within the CuO$_{2}$ planes, which are strongly
renormalized because of the strong interaction with
fluctuations of surrounding spinon excitations.

In the case of the incoherent charge dynamics in the c-axis
direction, {\it i.e.}, the independent electron propagation in each
layer, the c-axis holon current-current correlation function is then
proportional to the tunneling rate between just two adjacent planes,
and can be calculated in terms of the in-plane holon Green's
function $g_{ab}(k,\omega)$ by using standard formalisms for the
tunneling in metal-insulator-metal junctions \cite{n15,n18,n19} as,
\begin{eqnarray}
\Pi^{(h)}_{c}(i\omega_{n})=-(4\tilde{t}_{c}e\chi c_{0})^{2}{1\over N}
\sum_{k}{1\over\beta}\sum_{i\omega_{m}'}
g_{ab}(k,i\omega_{m}'+i\omega_{n})g_{ab}(k,i\omega_{m}'),
\end{eqnarray}
where $i\omega_{n}$ is the Matsubara frequency. Therefore the c-axis
current-current correlation function is essentially determined by
the property of the in-plane full holon propagator
$g_{ab}(k,i\omega_{n})$, which can be expressed as,
\begin{eqnarray}
g_{ab}(k,i\omega_{n})={1\over g^{(0)-1}_{ab}(k,i\omega_{n})-
\Sigma_{h}^{(2)}(k,i\omega_{n})}={1\over i\omega_{n}-\xi_{k}-
\Sigma_{h}^{(2)}(k,i\omega_{n})},
\end{eqnarray}
where the holon self-energy has been obtained by considering the
second-order correction due to the spinon pair bubble as \cite{n25},
\begin{eqnarray}
\Sigma_{h}^{(2)}(k,i\omega_{n})=(Zt)^{2}{1\over N^2}\sum_{pp'}
(\gamma_{p'-k}+\gamma_{p'+p+k})^{2}{B_{p'}B_{p+p'}\over
4\omega_{p'}\omega_{p+p'}}\times \nonumber \\
\left ( 2{n_{F}(\xi_{p+k})
[n_{B}(\omega_{p'})-n_{B}(\omega_{p+p'})]-n_{B}(\omega_{p+p'})
n_{B}(-\omega_{p'})\over i\omega_{n}+\omega_{p+p'}-\omega_{p'}-
\xi_{p+k}} \right. \nonumber \\
+{n_{F}(\xi_{p+k})[n_{B}(\omega_{p+p'})-n_{B}(-\omega_{p'})]+
n_{B}(\omega_{p'})n_{B}(\omega_{p+p'})\over i\omega_{n}+\omega_{p'}
+\omega_{p+p'}-\xi_{p+k}} \nonumber \\
\left. -{n_{F}(\xi_{p+k)}[n_{B}(\omega_{p+p'})
-n_{B}(-\omega_{p'})]-n_{B}(-\omega_{p'})n_{B}(-\omega_{p+p'})
\over i\omega_{n}-\omega_{p+p'}-\omega_{p'}-\xi_{p+k}}\right ),
\end{eqnarray}
with $n_{F}(\xi_{k})$ and $n_{B}(\omega_{k})$ are the Fermi and
Bose distribution functions, respectively. For the convenience in
the following discussions, the above full holon in-plane Green's
function $g_{ab}(k,i\omega_{n})$ also can be expressed as frequency
integrals in terms of the spectral representation as,
\begin{eqnarray}
g_{ab}(k,i\omega_{n})=\int_{-\infty}^{\infty}{d\omega\over 2\pi}
{A_{h}(k,\omega)\over i\omega_{n}-\omega},
\end{eqnarray}
with the in-plane holon spectral function
$A_{h}(k,\omega)=-2{\rm Im} g_{ab}(k,\omega)$. Then the c-axis
optical conductivity in the present theoretical framework is
expressed \cite{n18} as $\sigma_{c}(\omega)=-{\rm Im}\Pi^{(h)}_{c}
(\omega)/\omega$.

\section{Charge dynamics for the chain copper oxide materials}

We firstly consider the chain copper oxide materials. From the
experiments testing the c-axis charge dynamics \cite{n10}, it has
been shown that the presence of the rather conductive Cu-O chains
in the underdoped and optimally doped regimes can reduce the
blocking effect, and therefore the c-axis charge dynamics in this
system is effected by the same electron interaction as that in the
in-plane. In this case, we substitute Eq. (14) into Eq. (11), and
evaluate the frequency summations, then the c-axis optical
conductivity for the chain copper oxide materials can be obtained
as \cite{n15,n18,n19},
\begin{eqnarray}
\sigma_{c}(\omega)={1\over 2}(4\tilde{t}_{c}e\chi c_{0})^2
{1\over N}\sum_k\int^{\infty}_{-\infty}{d\omega'\over 2\pi}
A_{h}(k,\omega'+\omega)A_{h}(k,\omega'){n_{F}(\omega'+\omega)
-n_{F}(\omega') \over \omega},~~~~
\end{eqnarray}
where the in-plane momentum is conserved. This c-axis conductivity
$\sigma_{c}(\omega)$ has been calculated numerically, and the result
at the doping $\delta=0.12$ (solid line), $\delta=0.09$ (dashed line),
and $\delta=0.06$ (dot-dashed line) for the parameters $t/J=2.5$,
$\tilde{t}_{c}/t=0.04$, and $c_{0}/a_{0}=2.5$  at the temperature
$T$=0 is shown in Fig. 1, where the charge $e$ has been set as the
unit. From Fig. 1, we find that there is two bands in $\sigma_{c}
(\omega)$ separated at $\omega \sim 0.4t$, the higher-energy band,
corresponding to the "midinfrared band" in the in-plane optical
conductivity $\sigma_{ab}(\omega)$ \cite{n2,n3,n25}, shows a broad
peak at $\omega\sim 0.7t$, in particular, the weight of this band
is strongly doping dependent, and decreasing rapidly with dopings,
but the peak position does not appreciably shift to higher energies.
On the other hand, the transferred weight of the lower-energy band
forms a sharp peak at $\omega<0.4t$, which can be described formally
by the non-Drude formula, and our analysis indicates that this peak
decay is $\rightarrow 1/\omega$ at low energies as in the case of
$\sigma_{ab}(\omega)$ \cite{n2,n3,n25}. In comparison with
$\sigma_{ab}(\omega)$ \cite{n25}, the present result also shows that
the values of $\sigma_{c}(\omega)$ are by $2\sim 3$ orders of
magnitude smaller than those of $\sigma_{ab}(\omega)$ in the
corresponding energy range. The finite temperature behavior of
$\sigma_{c}(\omega)$ also has been discussed, and the result shows
that $\sigma_{c}(\omega)$ is temperature dependent, the
higher-energy band is severely suppressed with increasing
temperatures, and vanishes at higher temperatures. These results
are qualitatively consistent with the experimental results \cite{n10}
of the chain copper oxide materials and numerical simulations
\cite{n15}.

With the help of the c-axis conductivity, the c-axis resistivity
can be obtained as $\rho_{c}=1/\lim_{\omega\rightarrow 0}\sigma_{c}
(\omega)$, and the numerical result at the doping $\delta=0.12$
and $\delta=0.06$ for the parameters $t/J=2.5$, $\tilde{t}_{c}/t=
0.04$, and $c_{0}/a_{0}=2.5$ is shown in Fig. 2(a) and Fig. 2(b),
respectively. In the underdoped regime, the behavior of the
temperature dependence of $\rho_{c}(T)$ shows a crossover from the
high temperature metallic-like ($d\rho_{c}(T)/dT>0$) to the low
temperature semiconducting-like ($d\rho_{c}(T)/dT<0$), but the
metallic-like temperature dependence dominates over a wide
temperature range. Therefore in this case, there is a general trend
that the chain copper oxide materials show nonmetallic $\rho_{c}(T)$
in the underdoped regime at low temperatures. While in the optimally
doped regime, $\rho_{c}(T)$ is linear in temperatures, and shows
the metallic-like behavior for all temperatures. These results are
also qualitatively consistent with the experimental results of the
chain copper oxide materials \cite{n10} and numerical simulation
\cite{n15}.

\section{Charge dynamics for the no-chain copper oxide materials}

Now we turn to discuss the c-axis charge dynamics of the no-chain
copper oxide materials. It has been indicated from the experiments
\cite{n9} that for the no-chain copper oxide materials the doped
holes may introduce a disorder in between the CuO$_{2}$ planes,
contrary to the case of the chain copper oxide materials
\cite{n10}, where the increasing doping reduces the disorder in
between the CuO$_{2}$ planes due to the effect of the Cu-O chains.
Therefore for the no-chain copper oxide materials, the disorder
introduced by doped holes residing between the CuO$_{2}$ planes
modifies the interlayer hopping elements as the random matrix
elements. In this case, only the in-plane holon density of states
(DOS) $\Omega_{h}(\omega) =1/N\sum_{k}A_{h}(k,\omega)$ enters the
holon current-current correlation function as in disordered systems
\cite{n15,n29}, and after a similar discussion as in Sec. II, we
find that the corresponding momentum-nonconserving expression of
the c-axis conductivity $\sigma^{(n)}_{c}(\omega)$ for the no-chain
copper oxide materials is obtained by the replacement of the
in-plane holon spectral function $A_{h}(k,\omega)$ in Eq. (15)
with the in-plane holon DOS $\Omega_{h}(\omega)$ as \cite{n15,n29},
\begin{eqnarray}
\sigma^{(n)}_{c}(\omega)={1\over 2}(4\bar{t}_{c}e\chi c_{0})^2
\int^{\infty}_{-\infty}{d\omega'\over 2\pi}\Omega_{h}(\omega'+
\omega)\Omega_{h}(\omega'){n_{F}(\omega'+\omega)-n_{F}(\omega')
\over \omega},~~~~
\end{eqnarray}
where the $\bar{t}_{c}$ is some average of the random interlayer
hopping matrix elements $(\tilde{t}_{c})_{li}$. We have performed a
numerical calculation for this c-axis conductivity $\sigma^{(n)}_{c}
(\omega)$, and the result at the doping $\delta=0.10$ (solid line),
$\delta=0.08$ (dashed line), and $\delta=0.06$ (dash-dotted line)
for the parameters $t/J=2.5$, $\bar{t}_{c}/t=0.04$ with the
temperature $T=0$ is plotted in Fig. 3. This result shows that the
c-axis conductivity $\sigma^{(n)}_{c}(\omega)$ contains two bands,
the higher-energy band, corresponding to the midinfrared band in the
in-plane conductivity $\sigma_{ab}(\omega)$ \cite{n2,n3,n25}, shows
a broad peak at $\sim 0.3t$. The weight of this
band is increased with dopings, but the peak position does not
appreciably shift to lower energies. As a consequence of this
pinning of the transferred spectral weight, the weight of the
lower-energy band, corresponding to the non-Drude peak in
$\sigma_{ab}(\omega)$ \cite{n2,n3,n25}, is quite small and does
not form a well-defined peak at low energies in the underdoped
and optimally doped regimes. In this case, the conductivity
$\sigma^{(n)}_{c}(\omega)$ at low energies cannot be described by
the over-damped Drude-like formula even an $\omega$-dependence of
the in-plane holon scattering rate has been taken into
consideration. In comparison with the momentum-conserving
$\sigma_{c}(\omega)$, our result also shows that at low energies
the suppression of the momentum-nonconserving $\sigma^{(n)}_{c}
(\omega)$ is due to the disordered effect introduced by doped
holes residing between the CuO$_{2}$ planes. These results are
qualitatively consistent with the experimental results of the
no-chain copper oxide materials \cite{n9}.

For the further understanding the transport property of the no-chain
copper oxide materials, we have also performed the numerical
calculation for the c-axis resistivity $\rho^{(n)}_{c}=1/
\lim_{\omega\rightarrow 0}\sigma^{(n)}_{c}(\omega)$, and the
results at the doping $\delta=0.10$ and $\delta=0.06$ for the
parameters $t/J=2.5$, $\bar{t}_{c}/t=0.04$, and $c_{0}=2.5a_{0}$
are shown in Fig. 4(a) and Fig. 4(b), respectively. In accordance
with the c-axis conductivity $\sigma^{(n)}_{c}(\omega)$, the
behavior of the c-axis resistivity $\rho^{(n)}_{c}(T)$ in the
underdoped and optimally doped regimes is the semiconducting-like
at low temperatures, and metallic-like at higher temperatures. In
comparison with the in-plane resistivity $\rho_{ab}(T)$ \cite{n25},
it is shown that the values of the c-axis resistivity
$\rho^{(n)}_{c}(T)$ for the no-chain copper oxide materials are by
$3\sim 4$ orders of magnitude larger than these of the in-plane
resistivity $\rho_{ab}(T)$ in the corresponding energy range,
which are also qualitatively consistent with the experimental
results of the no-chain copper oxide materials \cite{n9}.

\section{Summary and discussions}

In the above discussions, the central concerns of the c-axis charge
dynamics in copper oxide materials are the two dimensionality of the
electron state and incoherent hopping between the CuO$_{2}$ planes,
and therefore the c-axis charge dynamics in the present fermion-spin
picture based on the charge-spin separation is mainly determined by
the in-plane charged holon fluctuation for the chain copper oxide
materials and the in-plane charged holon fluctuation incorporating
with the interlayer disorder for the no-chain copper oxide materials.
In comparison with the in-plane resistivity $\rho_{ab}(T)$ \cite{n25},
it is shown that the crossover to the semiconducting-like range in
$\rho_{c}(T)$ is obviously linked with the crossover from the
temperature linear to the nonlinear range in $\rho_{ab}(T)$,
{\it i.e.}, they should have a common origin.

In the fermion-spin theory \cite{n16}, the charge and spin degrees
of freedom of the physical electron are separated as the holon and
spinon, respectively. Although both holons and spinons contributed
to the charge and spin dynamics, it has been shown that the
scattering of spinons dominates the spin dynamics \cite{n30}, while
the results of the in-plane charge dynamics \cite{n25} and present
c-axis charge dynamics show that scattering of holons dominates the
charge dynamics, the two rates observed in the experiments \cite{n2}
are attributed to the scattering of two distinct excitations,
spinons and holons. It has been shown that an remarkable point of the
pseudogap is that it appears in both of spinon and holon excitations
\cite{n1,n2,n3,n4,n5,n6,n7,n8,n9,n10}. The present study
indicates that the observed crossovers of $\rho_{ab}$ and
$\rho_{c}$ for copper oxide materials seem to be connected with the
pseudogap in the in-plane charge holon excitations, which can be
understood from the physical property of the in-plane holon DOS. The
numerical result of the in-plane holon DOS $\Omega_{h}(\omega)$ at
the doping $\delta=0.06$, $\delta=0.12$, and $\delta=0.15$ for the
parameter $t/J=2.5$ at the temperature $T$=0 is shown in Fig. 5(a),
Fig. 5(b), and Fig. 5(c), respectively. For comparison, the
corresponding mean-field result (dashed line) is also shown in Fig. 5.
While the in-plane holon density of states $\Omega_{h}(\omega)$ in the
underdoping $\delta=0.06$ as a function of energy for the temperature
(a) $T=0$, (b) $T=0.01J$, and (c) $T=0.2J$ is plotted in Fig. 6.
From Fig. 5 and Fig. 6, we therefore find that the in-plane holon
DOS in MFA consists of the central part only, which comes from the
noninteracting particles as pointed in Ref. \cite{n28}. After
including fluctuations the central part is renormalized and two side
bands \cite{n28} and a V-shape holon pseudogap near the chemical
potential $\mu$ in the underdoped regime appear. But these two side
bands are almost doping and temperature independent, while the
V-shape holon pseudogap is doping and temperature dependent, and
grows monotonously as the doping $\delta$ decreases, and disappear
in the overdoped regime. Moreover, this holon pseudogap also
decreases with increasing temperatures, and vanishes at higher
temperatures. Since the full holon Green's function (then the holon
spectral function and DOS) is obtained by considering the
second-order correction due to the spinon pair bubble, then the holon
pseudogap is closely related to the spinon fluctuation. For small
dopings and lower temperatures, the holon kinetic energy is much
smaller than the magnetic energy, {\it i.e.}, $\delta t\ll J$, in
this case the magnetic fluctuation is strong enough to lead to the
holon pseudogap. This holon pseudogap would reduce the in-plane
holon scattering and thus is responsible for the metallic to
semiconducting crossover in the c-axis resistivity $\rho_{c}$ and
the deviation from the temperature linear behavior in the in-plane
resistivity $\rho_{ab}$ \cite{n25}. This holon pseudogap will also
lead to form the normal-state gap in the system, and the similar
result has been obtained from the doped {\it kagom\'e} and
triangular antiferromagnets \cite{n31}, where the strong quantum
fluctuation of spinons due to the geometric frustration leads to the
normal-state gap. With increasing temperatures or dopings, the
holon kinetic energy is increased, while the spinon magnetic
energy is decreased. In the region where the holon pseudogap
closes, at high temperatures or at higher doping levels, the
charged holon scattering would give rise to the temperature linear
in-plane resistivity as well as the metallic temperature dependence
of the c-axis resistivity. Our results also show that $\rho_{ab}(T)$
is only slightly affected by this holon pseudogap \cite{n25}, while
$\rho_{c}(T)$ is more sensitive to the underlying mechanism.

In summary, we have studied the c-axis charge dynamics of copper
oxide materials in the underdoped and optimally doped regimes
within the $t$-$J$ model by considering the incoherent interlayer
hopping. Our result shows the c-axis charge dynamics for the chain
copper oxide materials is mainly governed by the scattering from
the in-plane fluctuation, and the c-axis charge dynamics for the
no-chain copper oxide materials is dominated by the scattering
from the in-plane fluctuation incorporating with the interlayer
disorder, which would be suppressed when the holon pseudogap opens
at low temperatures and lower doping levels, leading to the
crossovers to the semiconducting-like range in the c-axis
resistivity $\rho_{c}(T)$ and the temperature linear to the
nonlinear range in the in-plane resistivity $\rho_{ab}(T)$.
Because copper oxide materials are very complex systems,
it is also possible that the actual c-axis conductivity may be a
linear combination of the momentum-conserving $\sigma_{c}(\omega)$
and momentum-nonconserving $\sigma^{(n)}_{c}(\omega)$, but we
believe that $\sigma_{c}(\omega)$ should be the major part of the
c-axis conductivity in the chain copper oxide materials, while
$\sigma^{(n)}_{c}$ should be the major part of the c-axis
conductivity in the no-chain copper oxide materials.

\acknowledgments
The authors would like to thank Professor Ru-Shan Han, Professor
H. Q. Lin, and Professor T. Xiang for helpful discussions. This work
was supported by the National Natural Science Foundation under Grant
No. 19774014 and the State Education Department of China through the
Foundation of Doctoral Training.

\begin{figure}
\caption{The c-axis optical conductivity $\sigma_{c}$ for the chain
copper oxide materials at the doping $\delta=0.12$ (solid line),
$\delta=0.09$ (dashed line), and $\delta=0.06$ (dot-dashed line) for
$t/J=2.5$, $\tilde{t}_{c}/t=0.04$, and $c_{0}/a_{0}=2.5$ with the
temperature $T=0$.}
\end{figure}

\begin{figure}
\caption{The c-axis electron resistivity $\rho_{c}$ for the chain
copper oxide materials at $t/J=2.5$, $\tilde{t}_{c}/t=0.04$, and
$c_{0}/a_{0}=2.5$ for (a) the doping $\delta=0.12$ and (b)
$\delta=0.06$.}
\end{figure}

\begin{figure}
\caption{The c-axis optical conductivity $\sigma^{(n)}_{c}$ for the
no-chain copper oxide materials at the doping $\delta=0.10$ (solid
line), $\delta=0.08$ (dashed line), and $\delta=0.06$ (dot-dashed
line) for $t/J=2.5$, $\bar{t}_{c}/t=0.04$, and $c_{0}/a_{0}=2.5$
with the temperature $T=0$.}
\end{figure}

\begin{figure}
\caption{The c-axis electron resistivity $\rho^{(n)}_{c}$ for the
no-chain copper oxide materials at $t/J=2.5$, $\bar{t}_{c}/t=0.04$,
and $c_{0}/a_{0}=2.5$ for (a) the doping $\delta=0.10$ and (b)
$\delta=0.06$.}
\end{figure}

\begin{figure}
\caption{The in-plane holon density of states at $t/J=2.5$ for (a)
the doping $\delta=0.06$, (b) $\delta=0.12$, and (c) $\delta=0.15$.
The dashed line is the result at the mean-field level.}
\end{figure}

\begin{figure}
\caption{The in-plane holon density of states for $t/J=2.5$ at the
doping $\delta=0.06$ for the temperature (a) $T=0$, (b) $T=0.01J$,
and (c) $T=0.2J$.}
\end{figure}

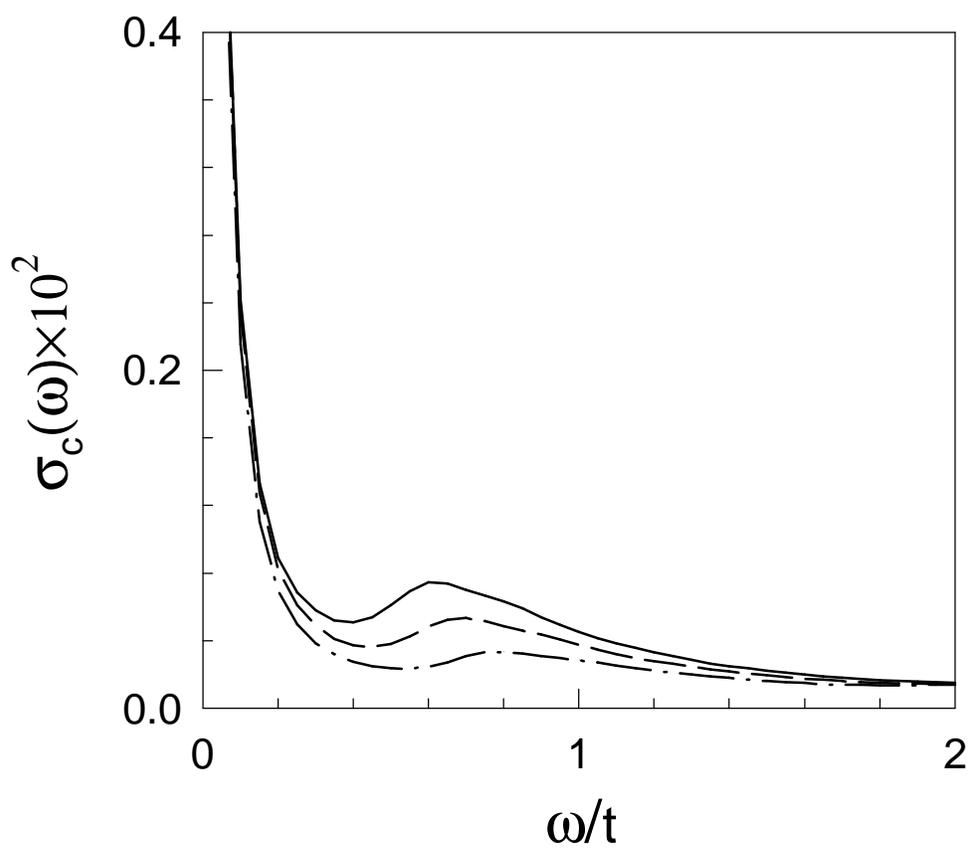
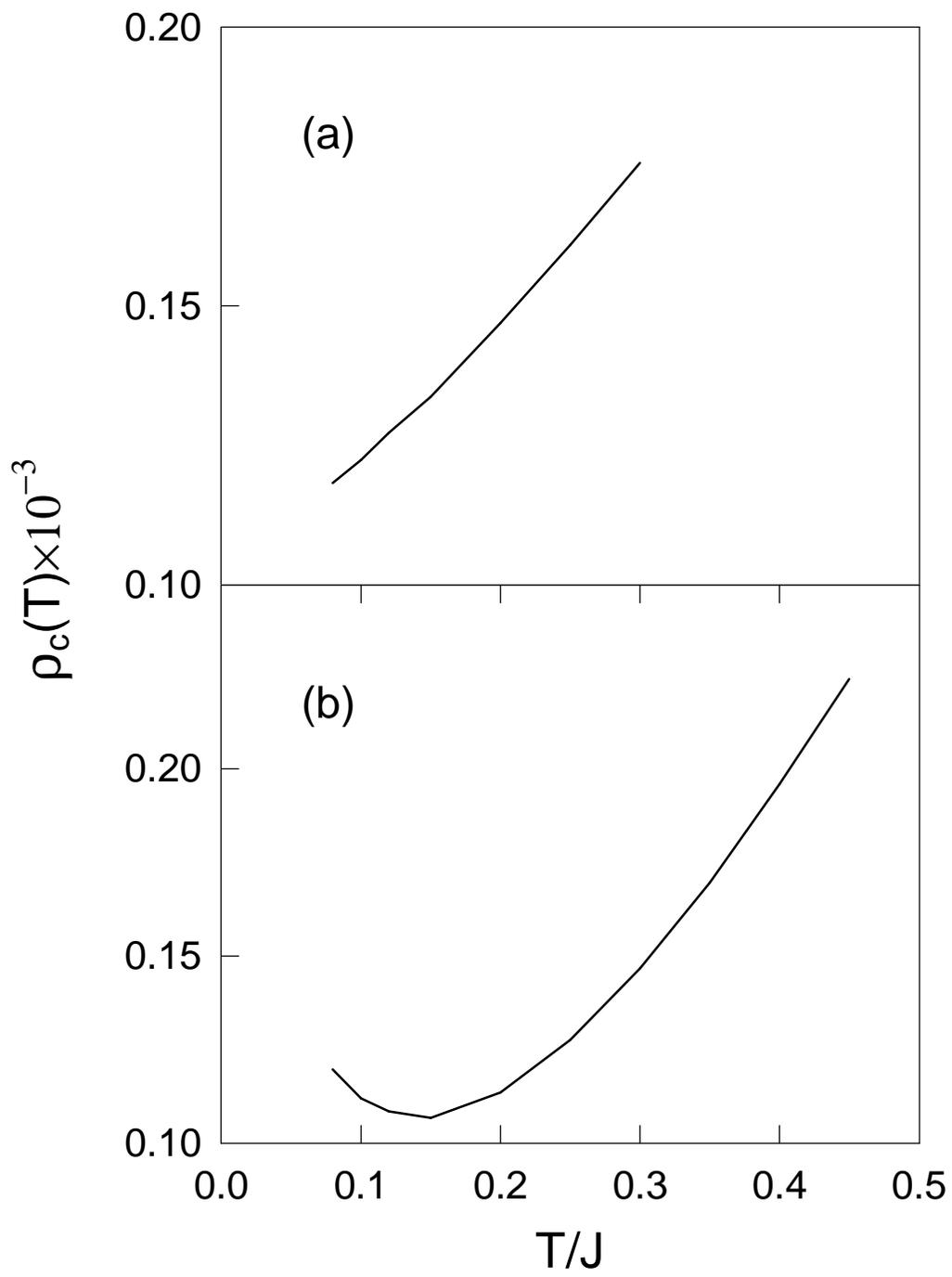
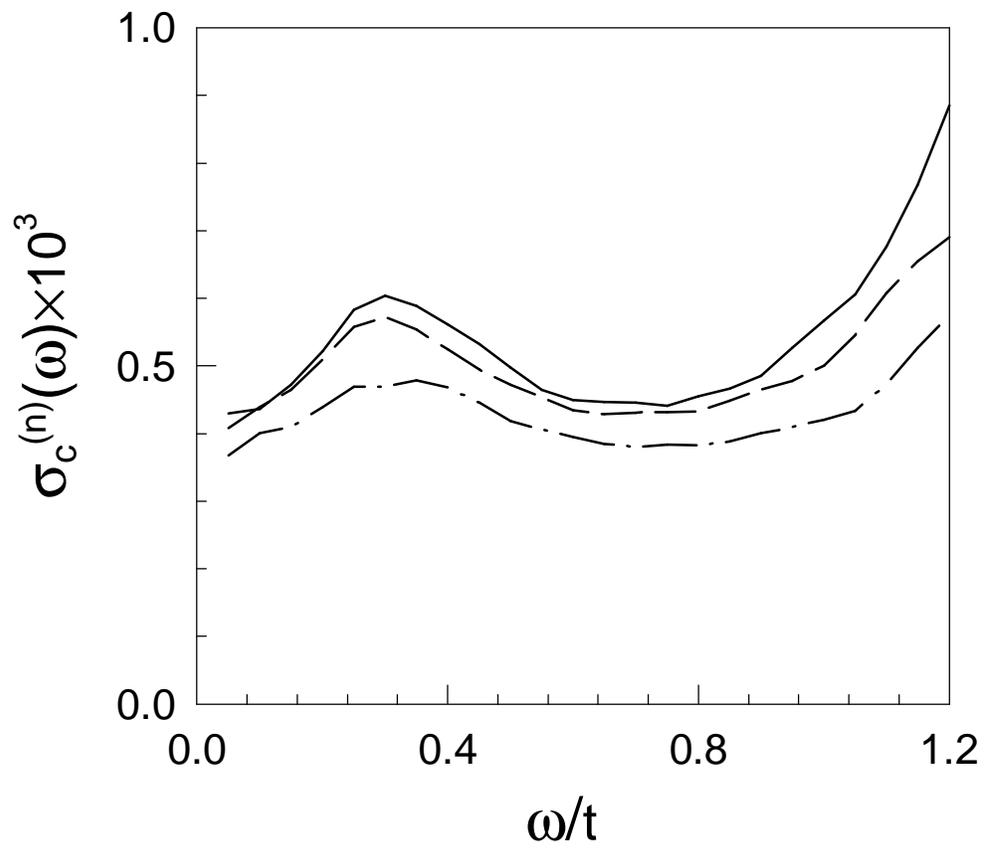
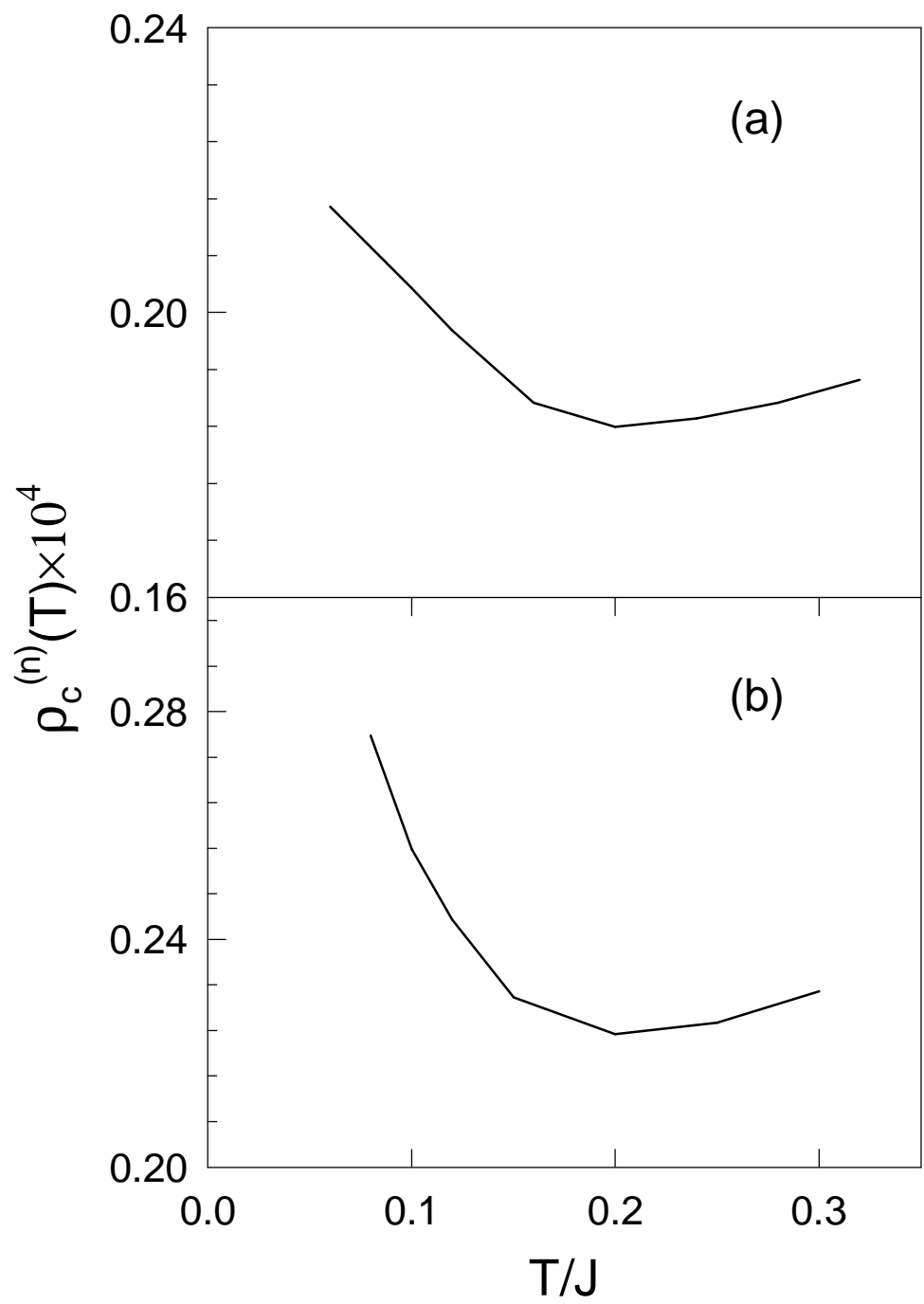
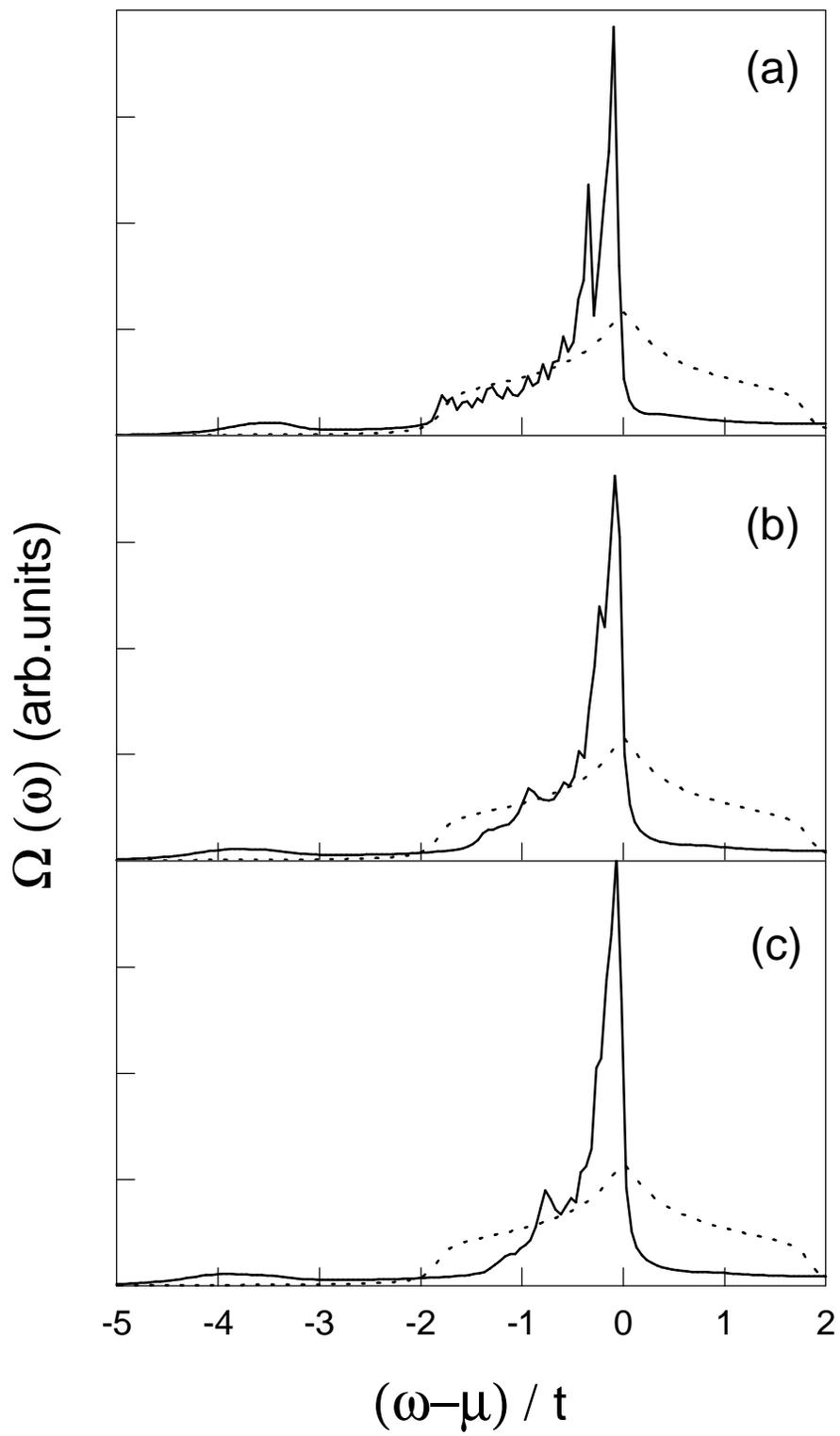
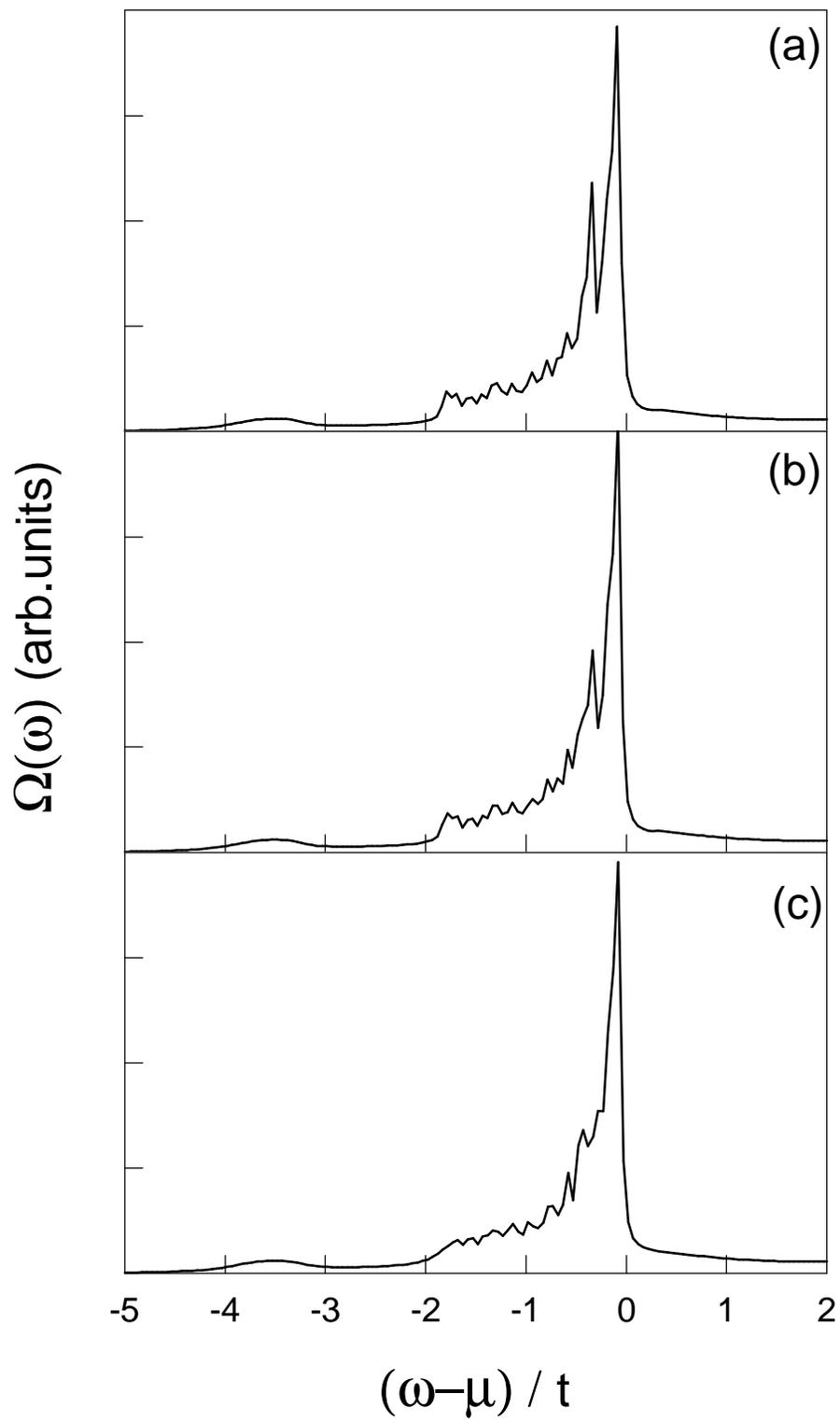
\end{document}